\documentclass{emulateapj}

\usepackage{graphicx}
\usepackage{psfig}

\slugcomment{ApJS in press}

\shorttitle{IRAC observations of the XFLS}
\shortauthors{Lacy et al.}

\begin{document}


\title{The Infrared Array Camera 
component of the {\em Spitzer Space Telescope} 
Extragalactic First Look Survey}


\author{M.\ Lacy\altaffilmark{1}, 
G.\ Wilson \altaffilmark{1}, F.\ Masci\altaffilmark{1}, 
L.J.\ Storrie-Lombardi\altaffilmark{1}, 
P.N.\ Appleton\altaffilmark{1}, L.\ Armus\altaffilmark{1}, 
S.C.\ Chapman\altaffilmark{1},
P.I.\ Choi\altaffilmark{1}, 
D. Fadda\altaffilmark{1},
F.\ Fang\altaffilmark{1},
D.T.\ Frayer\altaffilmark{1}, I.\ Heinrichsen\altaffilmark{1}, 
G.\ Helou\altaffilmark{1}, M.\ Im\altaffilmark{2}, S.\ Laine\altaffilmark{1}, 
F.R.\ Marleau\altaffilmark{1}, D.L.\ Shupe\altaffilmark{1}, 
B.T.\ Soifer\altaffilmark{1}, G.K.\ Squires\altaffilmark{1}, 
J.\ Surace\altaffilmark{1}, H.I.\ Teplitz\altaffilmark{1}, 
L.\ Yan\altaffilmark{1}} 
\altaffiltext{1}{Spitzer Science Center, Caltech, Mail Code 220-6,
Pasadena, CA 91125; mlacy@ipac.caltech.edu}
\altaffiltext{2}{Astronomy Program, School of Earth and Environmental 
Sciences, Seoul National University, Shillim-dong, Kwanak-gu, Seoul, S. Korea
2-880-9010}

\begin{abstract}
We present Infrared Array
Camera (IRAC) data and source catalogs from the Spitzer Space
Telescope Extragalactic First Look Survey.
The data were taken in four broad bands centered at nominal 
wavelengths of 3.6, 4.5, 5.8 and 8.0$\mu$m. 
A set of mosaics and catalogs have been produced  
which are $\approx 80$\% complete and $\approx 99$\%
reliable to their chosen flux density limits. The main field survey
covers $3.8 {\rm deg^{2}}$, and has flux density limits of 20$\mu$Jy,
25$\mu$Jy, 100$\mu$Jy and 100$\mu$Jy at wavelengths of 3.6, 4.5, 5.8 and 8.0$\mu$m, 
respectively. The
deeper ``verification'' survey covers $0.25 {\rm deg^{2}}$ with
limits of 10$\mu$Jy, 10$\mu$Jy, 30$\mu$Jy and 30$\mu$Jy, respectively. We also include deep data in the ELAIS-N1 field which 
covers $0.041 {\rm deg^{2}}$ with limits of 4$\mu$Jy, 3$\mu$Jy, 
10$\mu$Jy and 10$\mu$Jy, respectively,
but with only two wavelength coverage at a given sky position.
The final bandmerged catalogs contain 103193 objects in the 
main field, 12224 in the verification field and 5239 in ELAIS-N1. Flux 
densities of high signal-to-noise objects are accurate to about 10\%, and
the residual systematic error in the absolute flux 
density scale is $\sim 2-3$\%.
We have successfully extracted sources at source densities as high as 
100000 deg$^{-2}$ in our deepest 3.6 and 4.5$\mu$m data.
The mosaics and source catalogs will be made available through the Spitzer
Science Center archive and the Infrared Science Archive.
\end{abstract}



\section{Introduction}

The extragalactic portion of the {\em Spitzer Space Telescope} 
(Werner et al.\ 2004) First Look Survey
(hereafter XFLS) was one of the first observations made with {\em Spitzer} 
after the completion of Science Verification at the end of November 2003. 
The aim of this 67hr survey was to characterize the 
extragalactic source populations 
observed with {\em Spitzer} down to sub-milliJansky levels in the mid-infrared
\footnote{See the First Look Survey website at 
http://ssc.spitzer.caltech.edu/fls/}. Observations
covering the survey areas were made with both the Infrared
Array Camera (IRAC; Fazio et al.\ 2004a) and the Multiband Imaging 
Photometer for {\em Spitzer} 
(MIPS). This paper discusses the IRAC observations, 
the MIPS observations will be discussed in future papers
(Fadda et al.\ in preparation; Frayer et al.\ in preparation).
IRAC takes images in four broad
bands (termed channels). Channel 1 is centered at a nominal wavelength of 
 3.6$\mu$m, channel 2 at 4.5$\mu$m, channel 3 at 5.8$\mu$m and 
channel 4 at 8.0$\mu$m 
\footnote{The actual central wavelengths are slightly different and depend on the source spectrum, see Fazio et al.\ (2004) for
details.}. Light enters the instrument through
two apertures, one for channels 1 and 3, and one for channels 2 and 4. The
centers of the two apertures are separated by about $6\farcm 5$ on the sky.
Beamsplitters split the short and long wavelength light in each aperture.
The light in channels 1 and 2 is detected by two $256\times 256$ InSb arrays, and
that in channels 3 and 4 is detected by two $256 \times 256$  SiAs arrays. All
arrays have the same pixel size, corresponding to $1\farcs22$ per pixel, 
giving a $5\farcm 2 \times 5\farcm 2$ field of view. 
The pipeline-processed IRAC data from the XFLS were 
publicly released at the end of April 2004. In this paper, we 
describe the data analysis
which we performed to improve the data quality beyond the scope of the 
standard pipeline
processing, including a discussion of the removal of artifacts from the 
data. We present IRAC source catalogs for the XFLS, and discuss good 
observing practices for future surveys of this nature.

\section{Data Collection}

Nine Astronomical Observation Requests (AORs) cover the main field survey
area of 3.8deg$^{2}$ centered on R.A.(J2000) $17^{\rm h}18^{\rm m}00^{\rm s}$,
Dec.(J2000) $+59^{\circ}30^{'}00^{''}$ in a $3\times 3$ grid in array 
(row, column) coordinates. Each AOR was an approximately $8\times 8$ 
map with 277$^{''}$ offsets. The small 5-point Gaussian dither pattern with
12s frametime was 
used. This pattern has a mean offset of 28$^{''}$.
Data in the verification area (a 0.25 deg$^2$ area within the main field
centered on R.A.(J2000) $17^{\rm h}17^{\rm m}00^{\rm s}$, 
Dec.(J2000) $+59^{\circ}45^{'}00^{''}$) 
was taken using three AORs with 12s frametimes with the
same dither and mapping strategy as the main field 
and three much deeper AORs 
with 30s frametimes using 
the first sixteen points from the 
small cycling dither pattern, which has a mean offset of 13$^{''}$. 
These AORs are all contained
in {\em Spitzer} program identification number (PID) 26 (principal 
investigator (P.I.) T.\ Soifer).

In addition, a second field in the ELAIS-N1 region, 
centered on R.A.(J2000) $16^{\rm h}09^{\rm m}20^{\rm s}$,  
Dec.(2000) $+54^{\circ}57^{'}00^{''}$ was observed as part of 
a study of source confusion as a test field for the Great Observatories
Origins Deep Survey (GOODS), and this has also been included in the XFLS 
(PID 196, P.I.\ M.\ Dickinson). These 
data consisted of
a two position map separated by 312$^{''}$, with the first 36 dithers from
the medium cycling pattern and a 200s frametime. The map was oriented to 
avoid overlap between the channel 1/3 and channel 2/4 fields of view, thus, 
unlike the main and verification fields, a typical sky position in the image 
only has two channel coverage.

Details of the XFLS AORs, along with their AOR identification numbers 
(AORIDs) are given in Table 1. The AORs can be downloaded using the 
Spot tool 
(available from http://ssc.spitzer.caltech.edu/propkit/spot/), selecting
the ``View Program'' option and specifying the appropriate PID.

\begin{table*}

\caption{{\em Spitzer} observations of the XFLS fields}
{\scriptsize
\begin{tabular}{lcccccl}
AORID  &Date Obs& Field & Frametime & Map offset$^{*}$ &Mapping strategy$^{\dag}$ & Dither Pattern$^{\ddag}$\\
        &(UT)   &       &  (seconds) &(arcsec)         &                          &\\\hline
\dataset[ads/sa.spitzer/#0003861504]{3861504}&2003-12-01&main&12&-2273.5,-2550.5 &$7\times 9, \, (277^{''},277^{''})$  & 5-point Gaussian, small\\
\dataset[ads/sa.spitzer/#0003861760]{3861760}&2003-12-01&main&12&-2412.0,-138.5&$8\times 9, \, (277^{''},277^{''})$  & 5-point Gaussian, small\\
\dataset[ads/sa.spitzer/#0003862016]{3862016}&2003-12-02&main&12&-2412.0,2135.0&$8\times 8, \, (277^{''},277^{''})$  & 5-point Gaussian, small\\
\dataset[ads/sa.spitzer/#0003862272]{3862272}&2003-12-02&main&12&-138.5,-2412.0&$9\times 8, \, (277^{''},277^{''})$  & 5-point Gaussian, small\\
\dataset[ads/sa.spitzer/#0003862528]{3862528}&2003-12-03&main&12&-138.5,-138.5&$9\times 9, \, (277^{''},277^{''})$  & 5-point Gaussian, small\\
\dataset[ads/sa.spitzer/#0003862784]{3862784}&2003-12-03&main&12&-138.5,2273.5&$9\times 9, \, (277^{''},277^{''})$  & 5-point Gaussian, small\\
\dataset[ads/sa.spitzer/#0003863040]{3863040}&2003-12-03&main&12&2273.5,-2273.5&$9\times 7, \, (277^{''},277^{''})$  & 5-point Gaussian, small\\
\dataset[ads/sa.spitzer/#0003863296]{38633296}&2003-12-04&main&12&2273.5,-138.5&$9\times 9, \, (277^{''},277^{''})$  & 5-point Gaussian, small\\
\dataset[ads/sa.spitzer/#0003863552]{3863552}&2003-12-04&main&12&2135.0,2412.0&$8\times 10, \, (277^{''},277^{''})$  & 5-point Gaussian, small\\
\dataset[ads/sa.spitzer/#0003866880]{3866880}&2003-12-05&verification&12&0.0,0.0&$4\times 7, \, (277^{''},277^{''})$  & 5-point Gaussian, small\\
\dataset[ads/sa.spitzer/#0007676928]{7676928}&2003-12-05&verification&12&-830.0,-150.0&$2\times 6, \, (277^{''},277^{''})$  & 5-point Gaussian, small\\
\dataset[ads/sa.spitzer/#0007677184]{7677184}&2003-12-05&verification&12&830.0,150.0&$2\times 6, \, (277^{''},277^{''})$  & 5-point Gaussian, small\\
\dataset[ads/sa.spitzer/#0003867136]{3867136}&2003-12-05&verification&30&0,0&$4\times 5, \, (277^{''},277^{''})$  & Points 1-16 from small cycling\\
\dataset[ads/sa.spitzer/#0003867392]{3867392}&2003-12-05&verification&30&830.0,150.0&$2\times 5, \, (277^{''},277^{''})$  &Points 1-16 from small cycling\\
\dataset[ads/sa.spitzer/#0003867648]{3867648}&2003-12-06&verification&30&-830.0,-150.0&$2\times 5, \, (277^{''},277^{''})$  &Points 1-16 from small cycling\\
\dataset[ads/sa.spitzer/#0003867648]{6006016}&2003-12-28&ELAIS-N1&200&201.65,0.0&$1\times 2, \, (292^{''},312^{''})$  &Points 1-36 from medium cycling\\
                                             &&        &   &          &                                     &\\
\end{tabular} 

$^{*}$ relative to the appropriate field center, in array coordinates.

\noindent
$^{\dag}$ size of map (row$\times$column), mapping steps in array
coordinates.

\noindent
$^{\ddag}$ these are standard IRAC dither patterns, described in 
the 
{\em Spitzer} Observers' Manual http://ssc.spitzer.caltech.edu/documents/som.
}
\end{table*}   

\section{Data Analysis}

\subsection{Pipeline Processing}

The main and verification field data 
were run through the S10.5 version of the 
{\em Spitzer} Science Center (SSC) pipeline, described in 
the IRAC Data Handbook\footnote{Available from the SSC website 
(ssc.spitzer.caltech.edu/irac/dh/)}. Those for the ELAIS-N1 field used the 
S9.5 version of the pipeline. Between S9.5 and S10.5 the treatment of the
darks was improved, but the long frames in the ELAIS-N1 data are background
dominated, so the improvement between S9.5 and S10.5 was negligible for these
data. For each IRAC frame (termed a Data Collection Event (DCE)) the IRAC
pipeline produces processed images, the
Basic Calibrated Data (BCD), and corresponding 
masks (the DCE masks, or Dmasks) which we used as the starting point for 
our analysis. The masks flag potential problems with the data such as 
saturated pixels, strong radiation hits or corrupted/missing data.

\begin{figure*}

\plotone{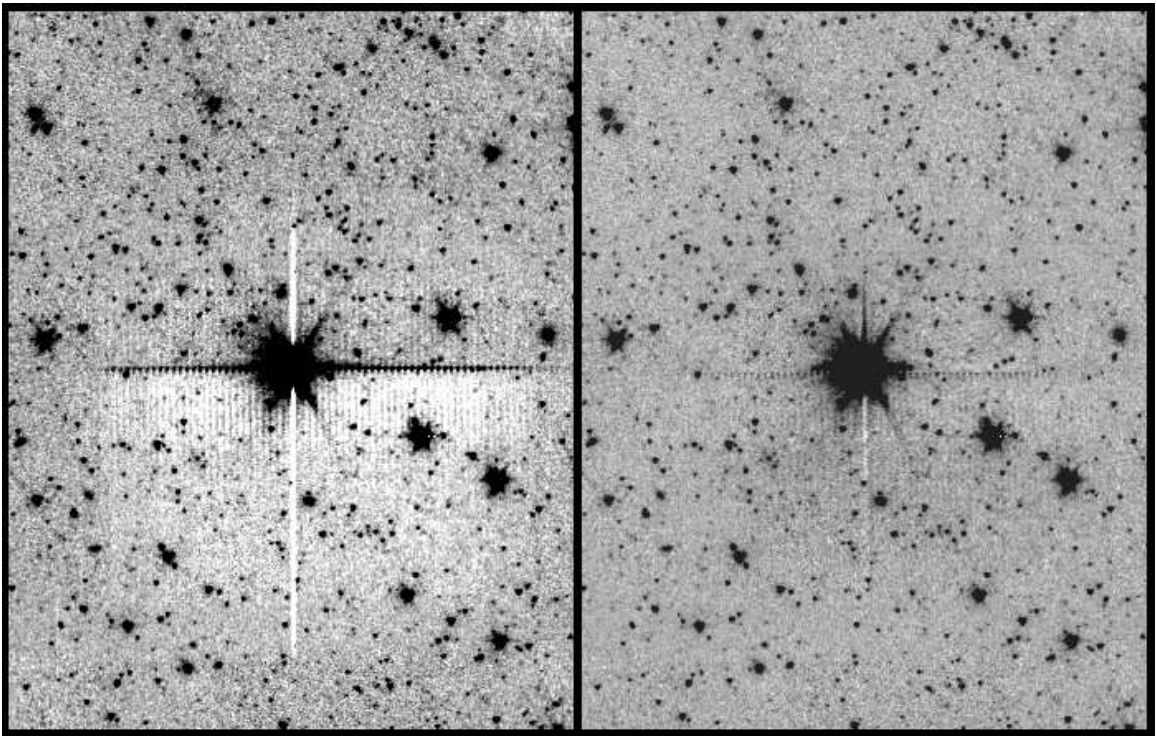}

\caption{A bright star in the channel 1 mosaic shown before (left) and 
after (right) the post-pipeline processing was applied. The star shows all the
artifacts produced by a bright star in channels 1 and 2: column pulldown,
the reduction in the data values in columns 
directly above and below the star,
muxbleed (raising the level of pixels in rows either side of the star), 
and the offset in the background
level produced by extremely strong muxbleed affecting all columns below the
star, and also producing striations in the background level. The post-pipeline
processing is able to remove most of the artifacts.}

\end{figure*}

\begin{figure*}
\plottwo{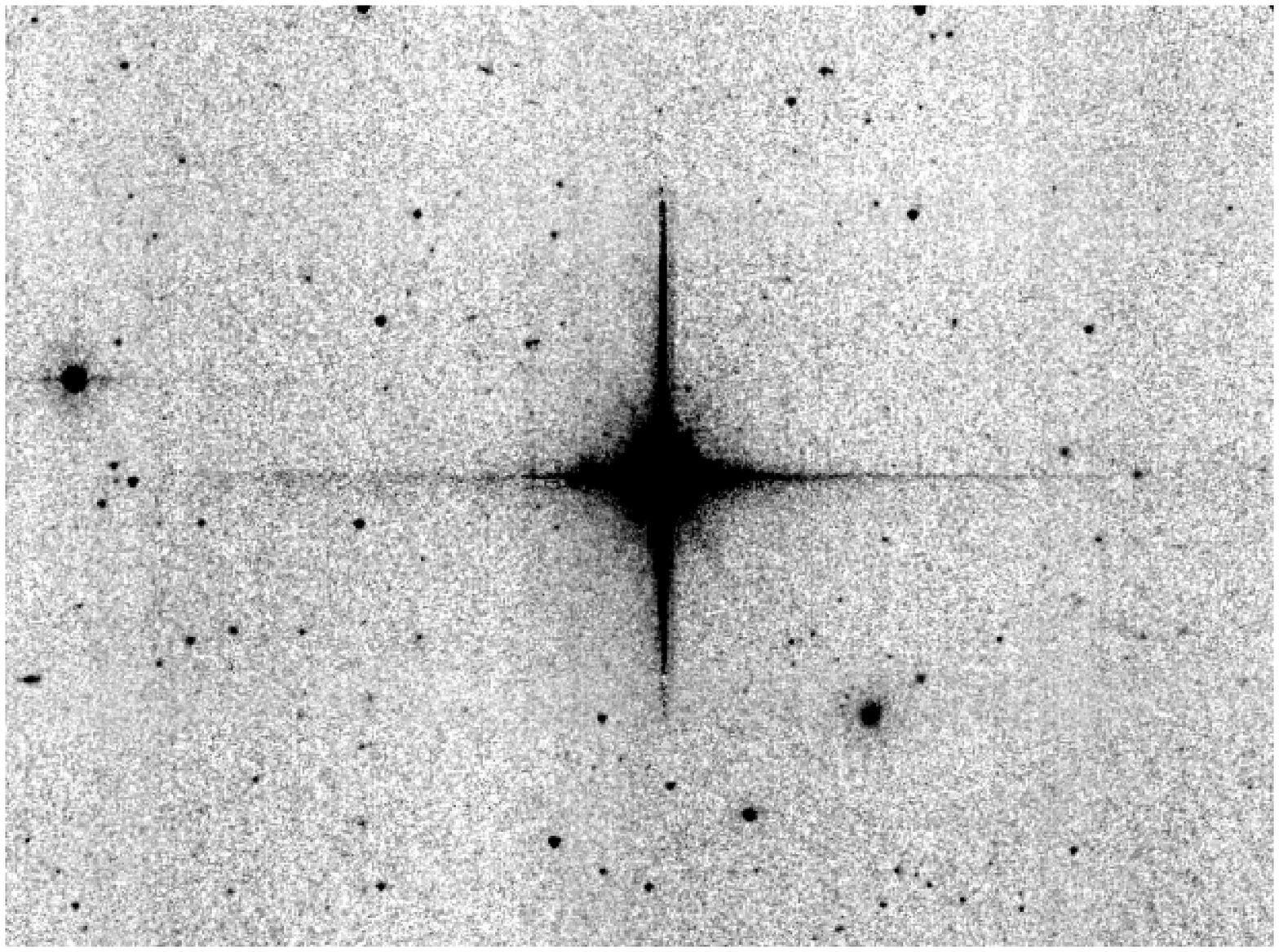}{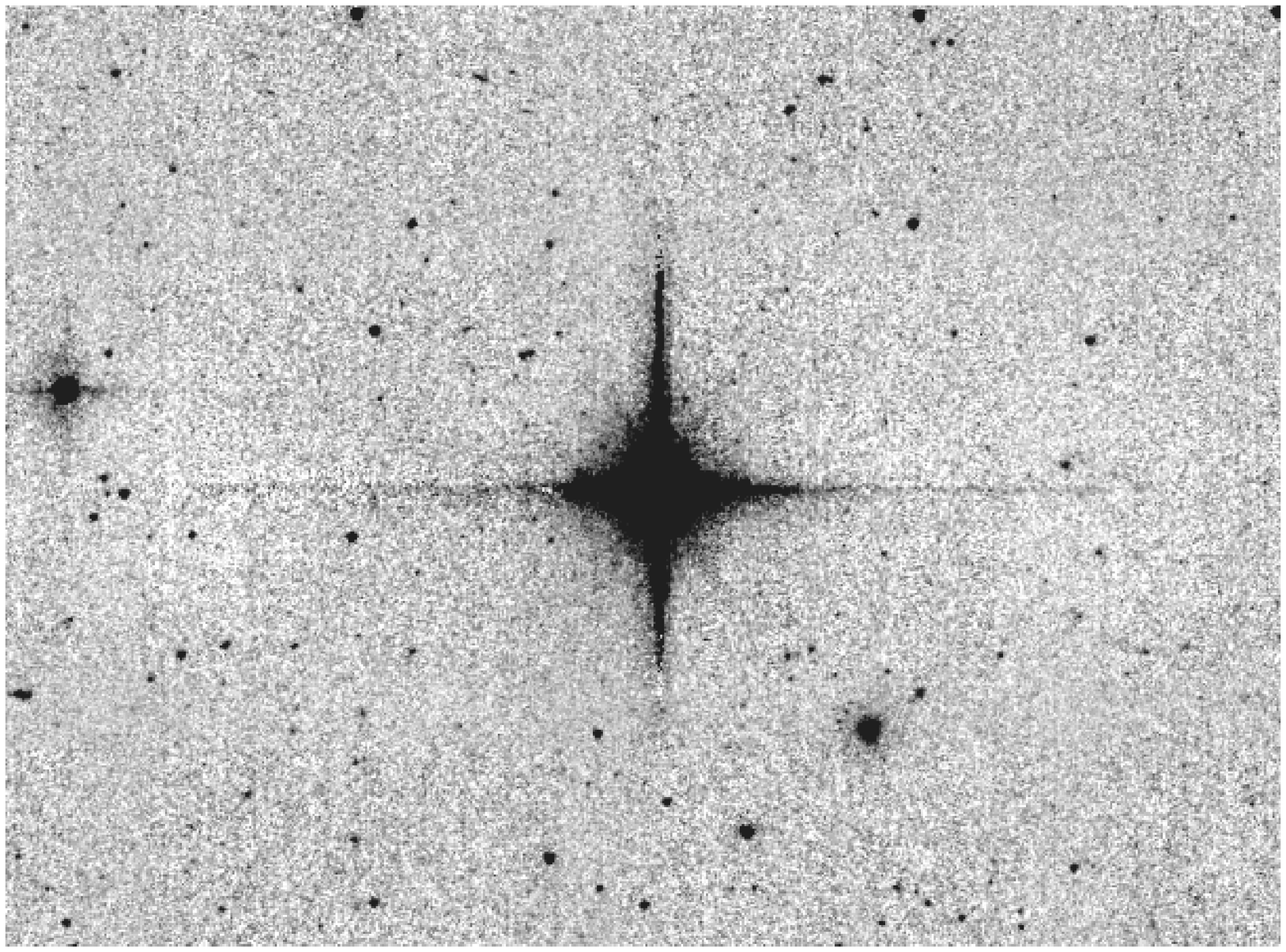}
\caption{A bright star in the channel 3 mosaic shown before (left) and 
after (right) the post-pipeline processing was applied. The star shows the
combination of optical banding and electronic 
artifacts produced by a bright star in channels 3 and 4, resulting
in the raised values of the pixels in the rows and columns on which 
a bright star falls. In the channel 3 data the artifacts are 
more pronounced in the column than the row direction, but in 
channel 4 the stronger artifacts are in the row direction. See the IRAC
Data Handbook for more details.}
\end{figure*}

\subsection{Post-pipeline Processing}

The IRAC data contain a number of artifacts which are not yet removed well by
the SSC pipeline. Some of these were removed in post-pipeline processing 
steps. 

Bright stars and cosmic rays 
systematically decrease the bias level in the columns in which 
they lie in channels 1 and 2 (Figure 1). This ``column
pulldown'' was removed using an algorithm 
developed and implemented by L.\ Moustakas and D.\ Stern
(personal communication). This masks objects in the frame to produce
a background image, then searches for 
discrepant columns, correcting them by adding a constant to bring the 
mean background value in the column up to that of its neighbors. 
Multiplexer bleed (``muxbleed'') in channels 1 and 2 from stars and cosmic 
rays with peak fluxes 
in the linear regime was partly corrected by the SSC pipeline, but that from 
the brightest, saturated stars was not. To correct these, we 
searched for bright stars in the frame, then split the data into the four 
amplifier outputs (every fourth column of data passes through the same output
amplifier). For each of the four outputs, we fit the rows within five pixels 
of the row containing the 
star with a straight line using columns outside $\pm 30$ pixels of 
the star, subtracted the fit, and reassembled the frame. 
In channels 3 and 4, banding effects (predominately 
optical in origin, produced by scattering of light within the detector) 
were present (Figure 2). 
These resulted in the rows and columns
containing the bright stars to be elevated above the background. The 
banding artifacts were removed in a similar fashion to the pulldown and
muxbleed in channels 1 and 2, though in this 
case the outputs were not split. Muxbleed from 
extremely bright stars in channels 1 and 2 can result in a large fraction
of the array being offset from the mean level. This  
was corrected by searching for saturated stars, then averaging the rows 
above and below the star in each of the four outputs, fitting a line to the 
mean of each row and subtracting the fit. 
Further details on and examples of these artifacts may be found in the 
IRAC Data Handbook.

The IRAC data have a number of problems which affect the background 
level across frames. 
The ``first frame effect'' - the variation in the dark which depends
on the length of time since the previous exposure - is partly corrected in 
the pipeline, but some residual variation in the dark remains, particularly
in channel 3. There are two reasons for this. First, the first frame
calibrations (taken prior to launch), did not cover a sufficiently large  
time range to calibrate the longest-term variations, 
and second, they were taken prior to the
decision to anneal the IRAC arrays regularly to remove long-term latents.
To remove the dark variations to first order,
a plane was fitted to produce a flat image with a constant background. 
This dealt with much of the uncorrected first frame effect. 
However, background variations remained from three sources which 
produce background variations on scales smaller than the array. These were:
residual first-frame/array relaxation effects, long-term latents 
(particularly noticeable in channel 1), and some real positional
variations in the background
level due to high Galactic latitude dust emission in channel 4. 
We therefore
subtracted a ``delta dark'' from each frame. In the data taken 
with the 12s frametimes, channels 1, 2 and 4 were adequately corrected 
(in the sense that the variations in the background level were reduced to 
be similar in magnitude to the noise level)
by subtracting a $\pm 1.5\sigma$-clipped median of the whole 
pixel stack from all 
the frames. Channel 3 required a running median of 15 of the 
$\pm 1.5\sigma$-clipped pixel stack to produce acceptable results. The
30s frametime data were more problematic.  
Channel 1 proved to be particularly challenging. 
The small dither pattern proved too small 
to effectively remove latent images from bright stars. In addition, the 
longer frametimes lead to a gradual build-up of latents over the whole array. 
A running median of 24 of the $\pm 1.5\sigma$-clipped pixel stack produced
good results for all but the last two pointings of the AORs (which happened
to contain bright stars). These frames had a 48-frame running median 
subtracted instead. In channel 3 the first 30s frame of the AOR 
was essentially uncorrectable
and was excluded from the mosaics. For the rest, a 36-frame running median
of the $\pm 1.5\sigma$-clipped pixel stack was found to be adequate. 
A straight median of the clipped frames, as used for the 12s data, 
was found to be adequate in channels 2 and 4.

To mask scattered light from sources off the detector array\footnote{For 
details of scattered light in IRAC see the {\em Spitzer} Observers' Manual http://ssc.spitzer.caltech.edu/documents/som/}, 
a straylight masking program developed by R.\ Arendt
and ML was used to place masks into the Dmask file. This program uses 
a list of bright stars from the Two Micron All Sky Survey (2MASS; 
Cutri et al.\ 2003)
to predict the positions of light scattered from the focal plane 
assembly (FPA) cover (the principal source of scattered light in channels 1 
and 2), 
and scattering off the edge of the detector (the principal source of 
scattering in channels 3 and 4). These masks have been applied to the 
final mosaics. 

Software for IRAC artifact correction and masking, including up to date
versions of much of the code described above may be found on the 
Spitzer contributed software website 
(http://ssc.spitzer.caltech.edu/archanaly/contributed/ browse.html).

The SSC pointing refinement module  (Masci, Makovoz \& Moshir 2004)
was run on all the channel 1 and channel 2 BCDs. This module
improves the pointing of the BCDs by both registering each BCD with 
respect to overlapping BCDs (``relative refinement'') and by registering
the result with stars from the 2MASS catalog (``absolute refinement''). 
To apply these corrections to the channel 3 and 4 BCDs, which have 
too few stars for a reliable absolute correction, we computed a 
weighted average of the channel 1 and channel 2 correction offsets,
applied these new offsets to all channel 1,2,3 and 4 BCDs and finally 
computed the refined pointing for all the BCDs. The mosaics constructed
using the refined pointings from the BCD are well-registered to the 
2MASS frame (see Section 5 below).

\begin{table*}
\caption{Values of the more important mosaicer and Sextractor parameters}
{\scriptsize
\begin{tabular}{llcccc}
Common to all:    &   &            &         &           &      \\
Program (module)& parameter & channel 1 & channel 2& channel 3 &channel 4\\\hline
Mopex (DETECT)      &Detection\_Max\_Area& 9&9&9&9\\
Mopex (DETECT)      &Detection\_Min\_Area& 0&0&0&0\\
Mopex (DETECT)      &Detection\_Threshold& 5&5&5&5\\
Mopex (MOSAICINT) & INTERP\_METHOD & 1 &1&1&1\\
Mopex (MOSAICDUALOUTLIER)& MIN\_OUTL\_IMAGE &2&2&2&2\\
Mopex (MOSAICDUALOUTLIER)& MIN\_OUTL\_FRAC &0.2&0.2&0.2&0.2\\
Mopex (MOSAICOUTLIER) &THRESH\_OPTION& 2& 2&2&2\\
Mopex (MOSAICOUTLIER) &BOTTOM\_THRESHOLD& 4& 4&4&4\\
Mopex (MOSAICOUTLIER) &TOP\_THRESHOLD& 4& 4&4&4\\
Mopex (MOSAICOUTLIER) &MIN\_PIX\_NUM& 3& 3&3&3\\
Mopex (MOSAICRMASK)   &MIN\_COVERAGE&3&3&3&3\\
Mopex (MOSAICRMASK)   &MAX\_COVERAGE&100&100&100&100\\
Sextractor  & DETECT\_MINAREA&3&3&3&3\\
Sextractor  & DETECT\_THRESH&2&2&2&2\\
Sextractor  & ANALYSIS\_THRESH&2&2&2&2\\
Sextractor  & FILTER&N&N&N&N\\
Sextractor  & DEBLEND\_NTHRESH&32&32&32&32\\
Sextractor  & DEBLEND\_MINCONT&0.0001&0.0001&0.0001&0.0001\\
Sextractor  & BACK\_SIZE&16&16&16&16\\
Sextractor  & BACK\_FILTERSIZE&3&3&3&3\\
Sextractor  & BACKPHOTO\_TYPE&LOCAL&LOCAL&LOCAL&LOCAL\\\hline
               &   &  & & &\\
Main field:     &   &            &         &           &      \\
Program& parameter & channel 1 & channel 2& channel 3 & channel 4\\\hline
Mopex  & MOSAIC\_PIXEL\_RATIO\_X& 1  & 1  & 1 & 1\\
Mopex  & MOSAIC\_PIXEL\_RATIO\_Y& 1  & 1  & 1 & 1\\
Mopex (MOSAICRMASK)&RM\_THRESH&0.8&0.8&0.05&0.05\\
Sextractor&SATUR\_LEVEL&325&463&2274&878\\
Sextractor&GAIN&309&280&67&195\\\hline
               &   &  & & &\\
Verification field:     &   &            &         &           &      \\
Program& parameter & channel 1 & channel 2& channel 3 & channel 4\\\hline
Mopex  & MOSAIC\_PIXEL\_RATIO\_X& 1  & 1  & 1 & 1\\
Mopex  & MOSAIC\_PIXEL\_RATIO\_Y& 1  & 1  & 1 & 1\\
Mopex (MOSAICRMASK)&RM\_THRESH&0.8&0.8&0.05&0.05\\
Sextractor&SATUR\_LEVEL&126&180&883&341\\
Sextractor&GAIN&786&721&67&172\\\hline
               &   &  & & &\\
ELAIS-N1 field:     &   &            &         &           &      \\
Program& parameter & channel 1 & channel 2& channel 3 & channel 4\\\hline
Mopex  & MOSAIC\_PIXEL\_RATIO\_X& 2  & 2  & 2 &2\\
Mopex  & MOSAIC\_PIXEL\_RATIO\_Y& 2  & 2  & 2 &2\\
Mopex (MOSAICRMASK)&RM\_THRESH&0.8&0.8&0.8&0.8\\
Sextractor&SATUR\_LEVEL&17.4&24.9&122&195\\
Sextractor&GAIN&5679&5210&1244&876\\\hline
\end{tabular}
}
\end{table*}

\subsection{Production of the Mosaics}

The SSC mosaicing software {\em Mopex}\footnote{Described in 
http://ssc.spitzer.caltech.edu/postbcd/doc/ mosaiker.pdf}
was used to mosaic the individual BCD images. {\em Mopex} consists 
of a set of modules combined in a {\sc perl} wrapper script. Which 
modules are used, and the parameters which control the modules, are
determined by namelist files. The namelists used were similar to the 
one given in the IRAC Data Handbook. 
Standard linear interpolation was used for all the mosaics. 
Two modes of outlier
rejection were employed, the ``dual outlier'' rejection, which uses
a combination of spatial and temporal criteria to identify outliers, and
a purely temporal (``multiframe'') technique which uses the pixel
stack at a given image position to identify outliers. The multiframe 
technique is the most effective when the coverage is high 
($\stackrel{>}{_{\sim}}$4), but the dual outlier method is better on
the lower-coverage regions on the edge of the mosaics, or in regions masked
by the Dmasks. The single frame outlier module, which relies purely on
a spatial criterion to determine outliers, was not used.
The diffuse cosmic rays in channels 3 and 
4 required that the masks be grown around detected outlier pixels in 
the lower coverage data. This 
was achieved in practice by setting the parameter $RM\_THRESH$ to a low
value (0.05) in the mosaicer channel 3 and 4 namelists for the main 
and verification fields.
Coverage maps for each mosaic were output by the mosaicer, and were checked
by eye for evidence of over-zealous outlier rejection. 
To help with deblending sources 
the ELAIS-N1 data were mosaiced
with a two-to-one pixel ratio to better sample the data. 
The main field and verification strip 
retained their original sampling.
The {\em Mopex} namelists used for the XFLS will be made
available on the XFLS website so others can reproduce our results should
they wish to do so. Some of the more important parameter choices are 
listed in Table 2.

\begin{table}
\caption{Aperture corrections}

\begin{tabular}{ccccc}
Aperture & \multicolumn{4}{c}{Correction applied}\\
         & channel 1&channel 2&channel 3&channel 4\\\hline
$6\farcs00$ &1.167 &1.213 &1.237& 1.466\\
$9\farcs26$ & 1.091&1.117 & 1.100 &1.165\\
$14\farcs86$&1.042 &1.048 &1.042 &1.066\\
$24\farcs4$ &1.000 & 1.000&1.000 & 1.000\\
\end{tabular}
\end{table}

\section{Production of the Source Catalogs}

Single-band catalogs were produced from each mosaic image using 
{\em Sextractor}
(Bertin \& Arnouts 1996). The coverage maps produced by {\em Mopex}
were used as weight images. We have optimized the 
photometry for the faintest objects. A background mesh size of 16 
pixels with a filter width of three was used, and the ``local'' background
(measured in a 24 pixel thick annulus) applied. Large, extended objects 
are thus likely to have incorrect fluxes in the catalog and should have their 
fluxes remeasured from the mosaics. Four fixed aperture fluxes 
(aperture diameters $6\farcs00$, $9\farcs26$,  $14\farcs86$ and  
$24\farcs40$) plus an isophotal flux were measured. The IRAC point spread
function 
has broad wings compared to typical ground-based data, and 
aperture corrections are significant (see the IRAC Data Handbook for 
further details). 
The largest aperture
diameter ($24\farcs40$) 
is the same diameter that the IRAC calibration stars are
measured in. The fluxes of the calibration stars in this aperture are 
considered the total fluxes in the S9.5 and S10.5 pipeline
processed data, so this aperture
requires no correction. The smaller apertures were picked to match those 
used by the Sloan Digital Sky Survey (SDSS).  For these, aperture corrections
were applied using a lookup table derived from measurements of bright stars
in the XFLS data, and checked for consistency with those derived by the IRAC
instrument team and those in the IRAC Data Handbook (see Table 3). 
Although these corrections are not quite appropriate
for the typical XFLS galaxy, which is slightly extended in IRAC images, our
simulations showed that the application of this correction nevertheless 
significantly improved the flux densities. One of the aperture fluxes
is taken as a ``best'' flux in the catalog. The ``best'' aperture was chosen 
by comparing the geometric mean radius of the isophote 
($r_m = \sqrt{A/\pi}$ where $A$ is the isophotal area) to each fixed aperture
diameter. This radius is compared to the radii of the four apertures
$r_1,r_2,r_3$ and $r_4$. If $r_m < 1.1 r_1$ then the ``best'' aperture
radius, $r_b$, is set to $r_1$, if $1.1r_1 < r_m \leq 1.1 r_2$ then 
$r_b=r_2$  etc. If $r_m \geq 1.1r_4$  then the isophotal flux is used as
the ``best'' flux. Although somewhat arbitrary, this procedure ensures
that an aperture appropriate to the isophotal size of the 
source is selected as ``best'', and 
thus reduces noise and confusion in the flux density measurement.

Flag images based on the results of the muxbleed, pulldown and banding 
correctors, latent image masking in the final mosaic (applied by hand) and 
haloes of bright stars (using sizes predicted from 2MASS K-magnitudes) 
were applied to the {\em Sextractor} catalogs. Objects whose isophotes cross 
non-zero pixels in the 
flag image have a value in the image flag field derived from OR-ing 
together the flag 
image pixels which fall 
within the object isophotes. Flag values and their meaning,
for both the flag values output by {\em Sextractor} and those added by 
applying the flag image, are listed in Table 4.

The four IRAC single-band catalogs were also merged into a single, 4-band
catalogue. The matching procedure began by going through each source in the
band 1 catalog and searching for a match within a radius of 1.5 
pixels in bands 2-4. Sources in band 2 unmatched to band 1 sources were then 
matched to bands 3 and 4, and finally sources in band 3 unmatched to bands
1 and 2 where then matched to band 4. The radius of 1.5 pixels was chosen
as a result of trying several different match radii between 0.5 and 2.5 
pixels, and examining the number of matches between each band. If the 
match radius is smaller than the mean position uncertainty due to random 
errors, the number of matches increases rapidly with increasing match radius.
This rate of increase falls off when the increase in the number
of matches is mostly due to chance coincidences. The optimum radius
is, of course, a function of band, but the similarity of the PSFs in the 
IRAC bands is enough that a single radius of 1.5 pixels is a 
good pick for all four bands. The fluxes are ``best'' fluxes as described
above with the aperture set to that of the shortest wavelength band in 
which the source is detected. The greater of the single-band
catalog limit, or a three sigma limit were placed on non-detections
(or negative aperture flux densities). The detection
flag field in the four-band catalog has bit 0 set for a channel 1 detection,
bit 1 for channel 2, bit 2 for channel 3, and bit 3 for channel 4.

\section{Positional Accuracy}

For 2MASS sources in the main field detected in all four bands above the 
IRAC catalog
flux limits (3223 sources; typical flux densities 
of $\approx 1$mJy in channel 1), 
the mean radial position error with respect to 
2MASS is $0.25^{''}$, with no measurable mean offset. 
This degrades at faint flux levels due to 
both noise and source confusion. Close to the flux limits of the catalogs
our simulations suggest the positional error is $\approx 1^{''}$.

\begin{figure*}

\plotone{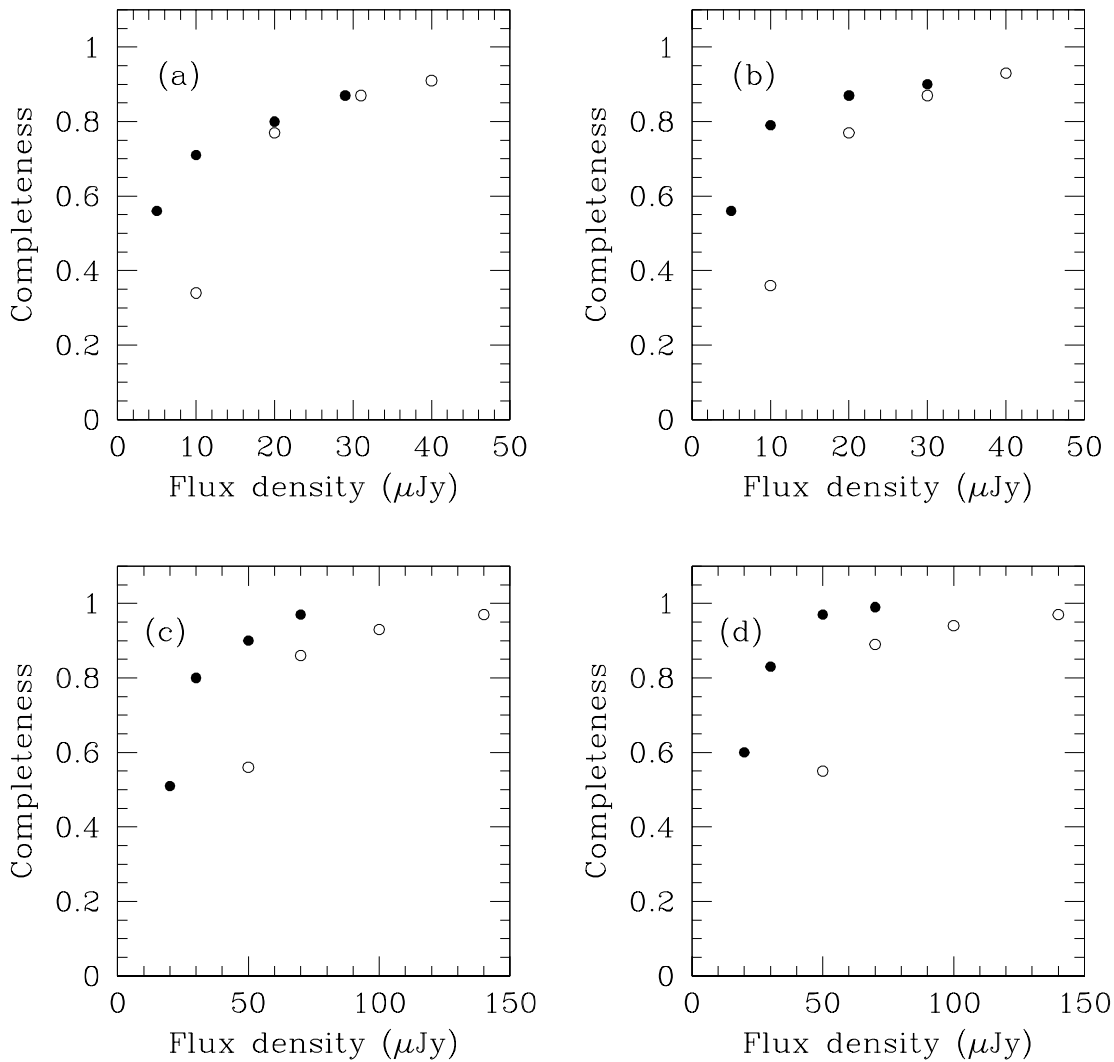}

\caption{Results of completeness tests on the main and verification
fields. The results for the main field are shown as open symbols, those 
for the verification field as closed symbols. The panels are for: (a) 
channel 1, (b) channel 2, (c) channel 3 and (d) channel 4.} 
\end{figure*}

\begin{figure*}

\plotone{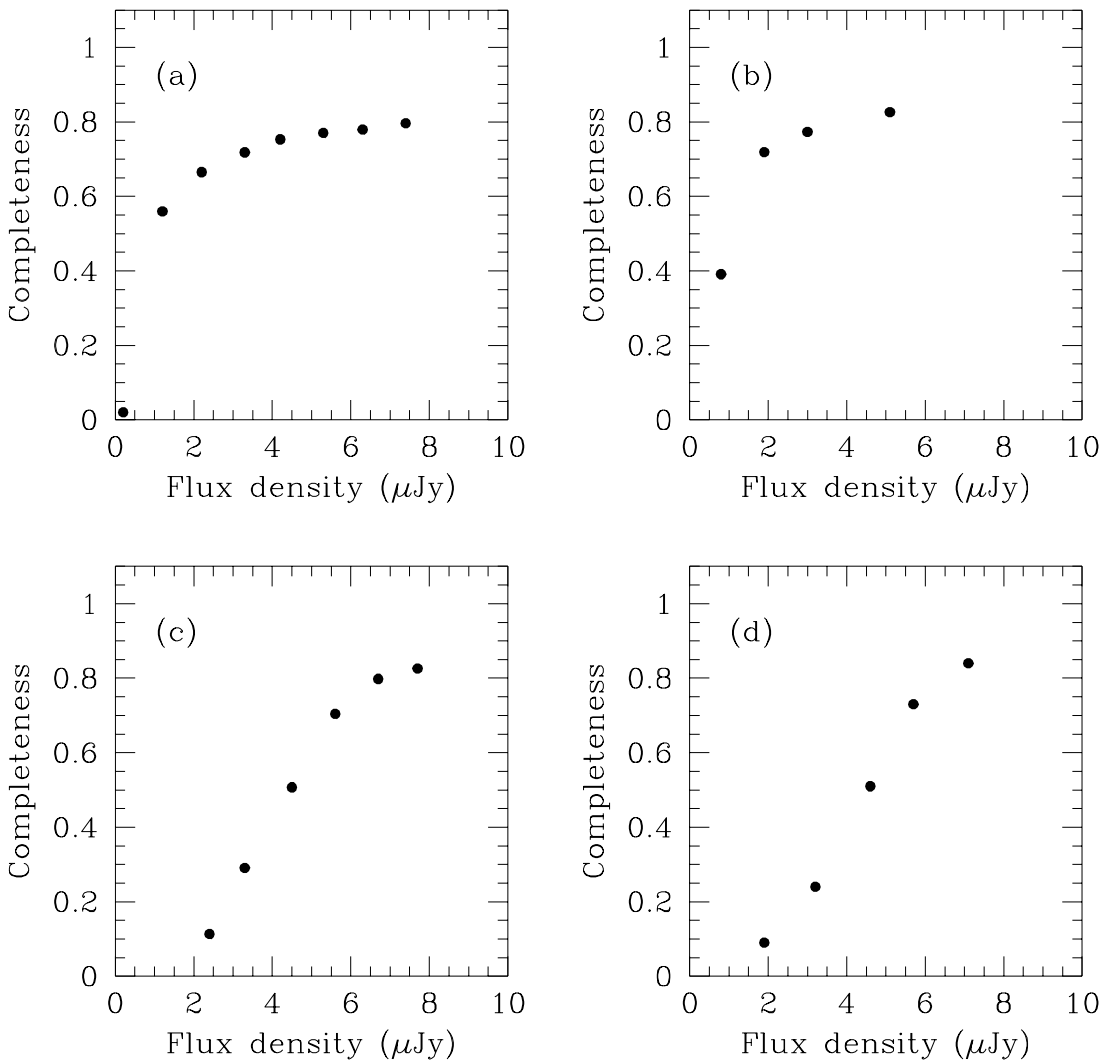}

\caption{Results of completeness tests on the ELAIS-N1
field. The panels are for: (a) 
channel 1, (b) channel 2, (c) channel 3 and (d) channel 4.} 
\end{figure*}

\begin{table}
\caption{Image flag values}

\begin{tabular}{cll}
Bit & meaning & set by\\\hline
0  & More than 10\% area affected by bad pixels&Sextractor\\
1  & Object originally blended with another one&Sextractor\\
2  & At least one object pixel is saturated&Sextractor\\
3  & Object is truncated at image boundary&Sextractor\\
4  & Object's aperture data are incomplete or corrupted&Sextractor\\
5  & Object's isophotal data are incomplete or corrupted&Sextractor\\
6  & A memory overflow occurred during deblending&Sextractor\\
7  & A memory overflow occurred during extraction&Sextractor\\
8  & Region affected by muxbleed, pulldown or banding& XFLS software\\
9  & Object in halo of bright star       & XFLS software\\
10 & Object contaminated by latent image & hand \\
\end{tabular}
\end{table}

\begin{table*}
\caption{Morphological breakdown for XFLS sources close to the survey limits}

\begin{tabular}{cccccccc}
Channel & flux range& \multicolumn{4}{c}{Fraction}  & Median $r_{1/2}$& range in $r_{1/2}$\\
        & ($\mu$Jy) & confused & point & disk & de Vaucouleurs & (arcsec)& (arcsec)\\\hline
1       & 20-30     & 0/10     & 2/10 & 6/10 & 2/10 & 0.4 & 0.2-0.7\\
1       & 10-20     & 0/10     & 3/10 & 7/10 & 0/10 & 0.4 & 0.3-1.1\\ 
1       &  3-4      & 1/10     & 0/10 & 8/10 & 1/10 & 0.07 & 0.03-0.18\\ 
4       &100-120    & 0/10     & 6/10 & 3/10 & 1/10 & 0.3  & 0.2-1.4\\ 
4       &30-40      & 1/10     & 3/10 & 5/10 & 1/10 & 0.4  & 0.2-0.8\\
4       &10-20      & 1/10     & 0/10 & 7/10 & 2/10 & 0.25 & 0.1-0.5\\
\end{tabular}

\end{table*}

\section{Completeness}

A series of simulations were run to estimate the completeness of the 
main, verification, and ELAIS-N1 surveys. We define completeness as
the chances of detecting a source placed at a random position in the 
field. Many sources are partially-resolved even at the limit of the surveys. 
To account for this in the completeness simulations we needed to obtain
an estimate of the range of source sizes near the survey limits. 
We used $i$-band {\em Hubble Space Telescope} images from the Advanced 
Camera for Surveys (ACS) taken as parallel data in program GO-9753 (P.I.\
Storrie-Lombardi) for which mosaics had been produced by I.\ Drozdovsky
(personal communication) to study the morphologies of 
objects in the main and verification fields.
For ELAIS-N1 no {\em HST} data were available, but the field is 
sufficiently deep that IRAC and ACS data from the 
Great Observatories Origins Deep Survey
(GOODS) in the Hubble Deep Field North (HDFN) could be used instead.
Ten galaxies close to the flux limits of the channel 1 and channel 4 
catalogs of each of the three surveys (main, 
verification and ELAIS-N1) were fit using software originally developed for 
quasar host galaxy fitting (Lacy et al.\ 2002).   
The results of this analysis are shown in Table 5. At the flux density
levels of the main and verification surveys, stars contribute $\sim 30$\% of 
the sources, but these disappear in the ELAIS-N1 survey. The median galaxy
half-light radius ($r_{1/2}$)
is 0.4$^{''}$ in the main and verification surveys falling to 
$<0.1^{''}$ in the channel 1 ELAIS-N1 data, and $0.25^{''}$ in the 
channel 4 ELAIS-N1 data. The distribution of galaxy types is 
field-galaxy like, with $\approx 20$\%
ellipticals, $\approx 80$\% spirals/irregulars. 

The scale sizes of sources in the main and verification fields mean that 
a significant fraction of these objects will be marginally resolved by 
Spitzer. In channel 1, the FWHM of the PSF is 1.8$^{''}$, similar to the FWHM
of an exponential disk galaxy with $r_{1/2}=0.5^{''}$. In the
ELAIS-N1 field, however, 
the scale sizes are sufficiently small that they can be 
treated as point sources for the purposes of completeness analysis. We 
therefore only modelled extended sources in the main and verification fields.
For modelling the completeness in these fields, 
artificial sources were added to 
the mosaics with fluxes, sizes and profiles which are
representative of the IRAC galaxy population close to the survey limit
based on the averaged results of our fitting. Specifically, we picked the 
model population as follows:
30\% point source, 60\% $r_{1/2}=0.4^{''}$ disk galaxies and 10\% 
$r_{1/2}=0.4^{''}$ de Vaucouleurs galaxies. For channels 3 and 4 close to the 
main field flux limit, we increased the fraction of point sources to reflect
the higher number of stars, using a mixture of 60\% point sources, 
30\%  $r_{1/2}=0.4^{''}$ disk galaxies and 10\% $r_{1/2}=0.4^{''}$
de Vaucouleurs galaxies.
We then estimated the completeness by 
comparing the number of these artificial objects of a given flux 
density appearing in the 
Sextractor catalogs with the number known to have been 
added to the mosaics. The completeness was not a strong function
of point versus extended source for the range of models used, or of
model type.
The completeness plots for the main and verification 
fields are shown in Figure 3. 
The ELAIS-N1 field has a very high 
source density in 
channels 1 and 2. For this field we adopted a slightly
different strategy by picking a typical field source from the mosaic, 
scaling it by different factors and inserting 1000 clones of this source
into the mosaic. The results are shown in Figure 4. 

We also tried using the verification data to check the completeness of the 
main field catalog, by examining the fraction of verification field sources
which were detected in the overlapping region of the 
main field catalog as a function of flux density 
close to the survey limits in the main field. 
This gave completeness values higher
by 10-15\% close to our adopted survey limits. This
result was not unexpected because a bright source will cause source 
confusion in both the main and the verification field, but provided a
useful check on the simulations.

We would like to emphasise that our completeness estimates are 
only approximate, particularly for the ELAIS-N1 channel
1 and 2 data, which is close to the confusion limit. A more careful
analysis of completeness in this confused regime (see, e.g., Chary et al.\
2004) will be necessary for accurate estimates of source counts at the
faintest flux density levels in these data.

\section{Photometric Accuracy}

An estimate of the uncertainty 
is associated with each flux density measurement in the catalog. This 
includes the uncertainties output by {\em Sextractor}, 
which are statistical in 
nature, and the known sources of systematic uncertainties. These have
been quantified by the IRAC Instrument and Instrument Support teams
(personal communications, though see discussions in the IRAC 
Data Handbook), and
are expected to amount to $\approx 10$\%. Most of this is due to 
color-dependence in the flat field, and is manifest as scatter in the 
cataloged fluxes. This value of 10\% 
was thus added in quadrature to the uncertainties from
{\em Sextractor}. We have used three approaches to checking the 
systematic photometric errors. The recovery of fluxes from artificial sources
was used to examine the uncertainties and biases in source measurement,
in particular those associated with
faint, slightly extended sources due to the small apertures and 
the consequent uncertainty in the aperture correction. Comparison was 
also made between fluxes measured for objects in the main field and the
(much deeper) verification field. Finally, comparison was made to 2MASS
photometry of field stars.

\subsection{Fluxes of Artificial Sources}

The catalog fluxes of the artificial
sources used in the completeness tests were compared
to their true fluxes. This tests the accuracy and 
effectiveness of our aperture 
corrections and photometric algorithms, 
but does not account for the color-dependent flat-field effect. 

These tests showed that, on average, more than 90\% of the flux is recovered
for point-like
or small sources close to the flux limit of the catalogs. Below the flux 
limits, the fraction of the model flux recovered falls fast. The most
likely reason for this bias is
source confusion and variations in the background level of the 
mosaic affecting the local background estimate used by {\em Sextractor}. 
For recovered objects which were not blended with bright neighbors the 
difference between their true and recovered fluxes was in line with 
the expected statistical error.

\subsection{Comparison Between the Main and Verification Surveys}

Fluxes of objects in the part of the main survey which overlapped with the 
verification strip were compared. This comparison allows a rough estimate
of the color-dependent flat field error, as the main and verification surveys
were taken with different grids, and of errors induced through over-zealous
cosmic ray rejection in the shallower coverage main field data. 

The results of this test showed that the scatter between the 
main field and verification strip fluxes is about 10\% in all four 
bands for high signal-to-noise detections, roughly as expected from the 
known sources of systematic error. To
get a true estimate of the error would require a more random superposition
of the main and verification survey grids, thus
this estimate (which corresponds to an $\approx 7$\% uncertainty 
if the main and verification regions are affected similarly) 
is strictly a lower limit. 

The main field fluxes in all bands
average a little lower than those in the verification
survey, by about 2\%. 
This is probably due to a combination of factors. At the faint end,
the fluxes are measured in small apertures in the shallower survey, 
frequently missing flux. At the bright end, the lower coverage of the main
survey will lead to more frequent mis-identification of the brightest
object pixels as cosmic rays, and the consequent exclusion of the 
brightest object pixels from the final mosaics. 

\subsection{Colors of 2MASS Stars}

We performed a further check on the photometry by studying the colors of 
field 2MASS stars in the XFLS main field, 
following Eisenhardt et al.\ (2004). 
We placed a color cut of $J-K < 0.3$ on the 2MASS stars, corresponding 
approximately to the color of a F7 dwarf (Bessell \& Brett 1988), and 
only considered stars with $K<14$ to ensure accurate 2MASS colors. Also,
we excluded stars with $K<10$ from the quantitative comparison in channel 
1 as these were saturated in the IRAC images. 
For channels 2, 3 and 4 the mean stellar color is close to zero, as expected.
In channel 1, our initial calibration showed a significant 
offset from zero, similar to that noted 
by Eisenhardt et al.. Part of this is due to the expected
non-zero colors of the stars. Most of the stars are close to the red limit
in $J-K$, which corresponds to a $K-L \approx K-{ch1}$ 
color of $\approx 0.04$ (see Table 6). However, 
a significant offset $\sim 5$\%-$8$\% still remained.  

Based on these plots, we decided to change the flux conversion factor 
for channel 1 to the value
determined from measuring the fluxes of the A-star calibrators only
(T.\ Megeath, personal communication), 0.1085,  4\% different from that
given in the S9.5/S10.5 BCD headers (0.1125). 
This improves the channel 1 fluxes further,
and the residual discrepancy is $\sim 3$\%. This 
residual is comparable to
that expected from the effects described above, and to the known 
uncertainties in the channel 1 flat fields from IRAC campaigns 1 and 2. 
Figure 5 shows the final plots, and Table 6 the numerical 
results. Based on these, the flux density scale of channel 1 looks accurate
to $\approx 3$\% both when compared to $K$-band (panel (b)) and to channel 2
(panel (c)). The apparent upswing in the mean color in the $K-ch2$ plot for 
$K>13$ (panel (d)) largely disappears
when a redder color cut (which includes more objects) is used, so 
is probably not significant. Examination of Table 6 suggests that, indeed,
our estimate of 10\% systematic uncertainty in the flux densities of
individual sources is reasonable, and that the systematic uncertainty
in the flux density scale is $\sim 2-3$\%. Unfortunately there were too few
stars in the verification region to allow us to perform the same test on 
the deeper data, but the comparison between the main and verification field
flux densities showed only small systematic differences (Section 7.2).

\begin{table}
\caption{Offsets and dispersion of the colors of blue stars between 2MASS and 
IRAC}
\begin{tabular}{cccc}
Color   &Expected  & Offset from    & Dispersion around\\ 
        &mean value$^{*}$& expected mean & expected mean\\\hline
$K-ch1$ & 0.04     & +0.03 & 0.07\\
$K-ch2$ & 0.02     & +0.00 & 0.12\\
$K-ch3$ & 0.00     & -0.02 & 0.11\\
$K-ch4$ & 0.00     & +0.02 & 0.12\\
\end{tabular}

\noindent 
$^{*}$ from Bessell and Brett (1988).

\end{table}

\section{Setting the Flux Density Limits}

As this survey is expected to be used primarily for statistical purposes, 
we have produced single-band catalogs cut at flux density levels 
at which the surveys are still $\approx 80$\% complete (Table 7), 
and at which the simulations indicate that the mean fraction of flux 
recovered by {\em Sextractor} 
from point or slightly extended sources is $\stackrel{>}{_{\sim}}90$\%.
This resulted in us setting limits in the main field catalog
of 20$\mu$Jy, 25$\mu$Jy, 100$\mu$Jy and 100$\mu$Jy in channels 1, 2, 3 and 4, 
respectively. The corresponding limits in the verification survey were 
10$\mu$Jy, 10$\mu$Jy, 30$\mu$Jy and 30$\mu$Jy, and those in the ELAIS-N1
field are 4$\mu$Jy, 3$\mu$Jy, 10$\mu$Jy and 10$\mu$Jy. The survey limits 
scale approximately with the square root of the exposure time in channels 
2--4, but channel 1 is less deep than expected in the verification and 
ELAIS-N1 fields. Source confusion is certainly partly to blame, but in 
channel 2 similar source densities are reached in the deep surveys without
a strong departure from the expected scaling. The most likely explanation
is a combination of source confusion and the latent image problem,
which was especially severe in the verification data in channel 1, and 
resulted in a varying background which was only partly corrected 
by the subtraction of a delta-dark. 
The flux density limits correspond to a signal-to-noise of $\approx 5$
for a typical source at the limit of the surveys (Figure 6), except in the
deeper channel 1 data affected by variable background and confusion, where 
it is $\approx 7$.

We have 
compared our flux density limits with those of the Spitzer Performance 
Estimation Tool (PET, http://ssc.spitzer.caltech.edu/tools/pet.html).
For the main field, we expect 5$\sigma$ sensitivities of 7.4, 11, 61 and
60$\mu$Jy in channels 1-4 respectively, compared to our survey limits
of 20, 25, 100 and 100$\mu$Jy, respectively. In the verification survey, 
the PET gives 5$\sigma$ sensitivities of 1.8, 3, 20 and 23 $\mu$Jy compared
to 10,10,30 and 30 $\mu$Jy, and in ELAIS-N1, 0.33, 0.7, 4.6 and 5.8 $\mu$Jy
compared with 4, 3, 10 and 10$\mu$Jy. There are three reasons why we
do not achieve the PET sensitivities. First, the PET estimates are based
on point-source fitting rather than aperture photometry. Point-source
fitting has 7-13 noise pixels, depending on channel 
(see the {\em Spitzer} Observers' 
Manual), compared to 19 pixels in our smallest photometric aperture, which
accounts for much of the difference in the main field
survey, and in all surveys in channels 3 and 4. Second, the 
PET estimates take no account of source confusion. This is particularly 
important for channels 1 and 2 in the ELAIS-N1 survey. Third, the background
variations in channel 1 discussed above limit our sensitivity in 
the deeper surveys in that channel.

\section{Reliability}

Objects detected 
in the main field catalogs which overlapped the region covered
by the verification data were compared with those made from the
30s data in the verification strip, which goes much deeper. This allowed
us to make an estimate of the reliability of the main field catalog at the
catalog flux density limits given above. These
tests showed that the catalogs are $\approx 99$\% 
reliable in all four bands, in the sense that the probability of 
an object in the main field catalog being detected in the deeper verification
data was $\approx 99$\%. Measuring
the reliability of the verification and ELAIS-N1 catalogs is not possible
using this technique, 
but as their limits are similar to the main field in terms of signal-to-noise
ratio we expect these catalogs to be similarly reliable.

\section{Data Products}

\subsection{Mosaics}

A mosaic for each channel for the main and verification fields is available,
along with coverage maps. All are made in the same fiducial frame, i.e.\ with
the same pixel grid.  This results in 
much of the verification mosaic being blank, but which greatly eases source 
comparison as the same pixels in every mosaic and in every channel correspond
to the same sky position. The 12s and 30s frames in the verification region
are coadded according to inverse variance weights determined from the 
measured standard deviations in the BCDs.
We opted to keep the main field data separate from the verification field
data to maintain their statistical independence. 
The coverage maps show the numbers of frames 
added to produce the final mosaic. In the main field, one unit of coverage
corresponds to one 12s frame, in the verification field one unit is one
30s frame, with the 12s frames being 0.388 (the ratio of the exposure
times in the 12s and 30s data). Multiplying by the appropriate
exposure time (10.4s for
the 12s frames in the mainfield and 26.4s for the 30s frames in the 
verification field) will thus recover the exposure
time for a given pixel.

\begin{figure*}

\plotone{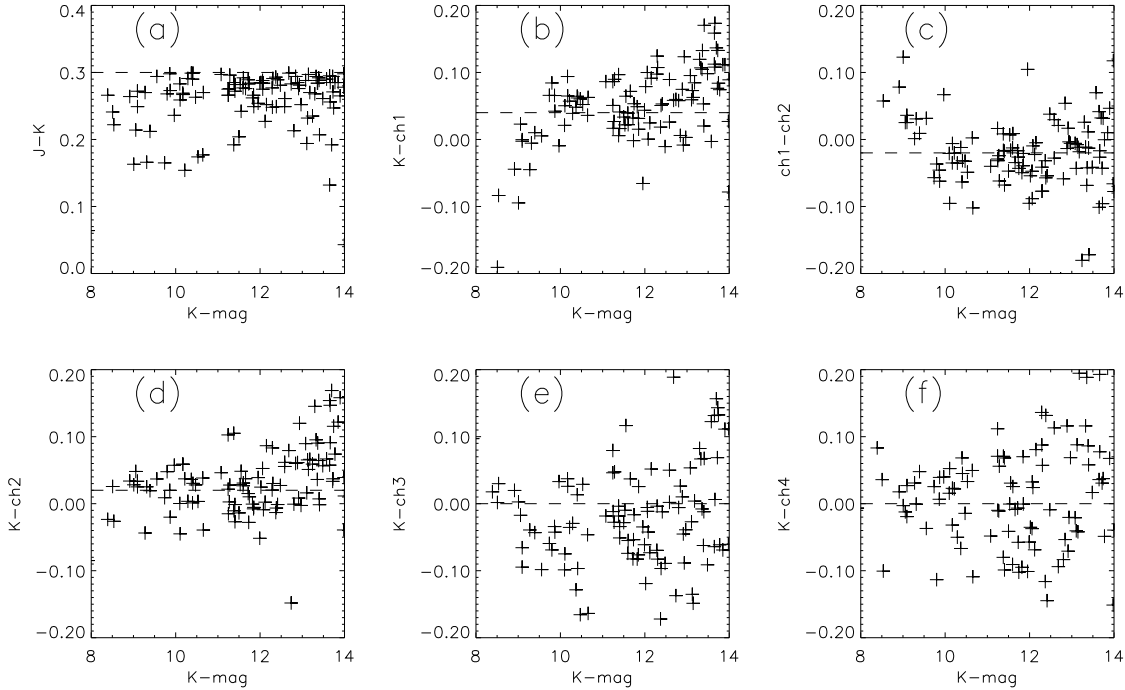}

\caption{Photometry of blue stars in the XFLS field. (a) $J-K$ versus
$K$ color-magnitude diagram. The dashed line shows the $J-K=0.3$ color
cut used to select the blue stars (corresponding to late-F and bluer). 
(b) $K-ch1$ vs $K$. The dashed line at $K- ch1=0.04$ 
is the expected 
color of stars with  $J-K=0.3$. (Stars brighter than $K\approx 10$ are
saturated in the channel 1 image, producing the downturn at bright magnitudes.)
(c) $ch1-ch2$ vs $K$. The dashed line
at $ch1-ch2=-0.02$ is the expected color of stars with  $J-K=0.3$. (d)
$K-ch2$ vs $K$. The dashed line
at $K-ch2=0.02$ is the expected color of stars with  $J-K=0.3$ , (e) $K-ch3$ vs $K$, and (f) 
$K-ch4$ vs $K$. In (e) 
and (f) the dashed line is at zero, the expected color in these
pairs of bands.}

\end{figure*}

\begin{figure*}

\plotone{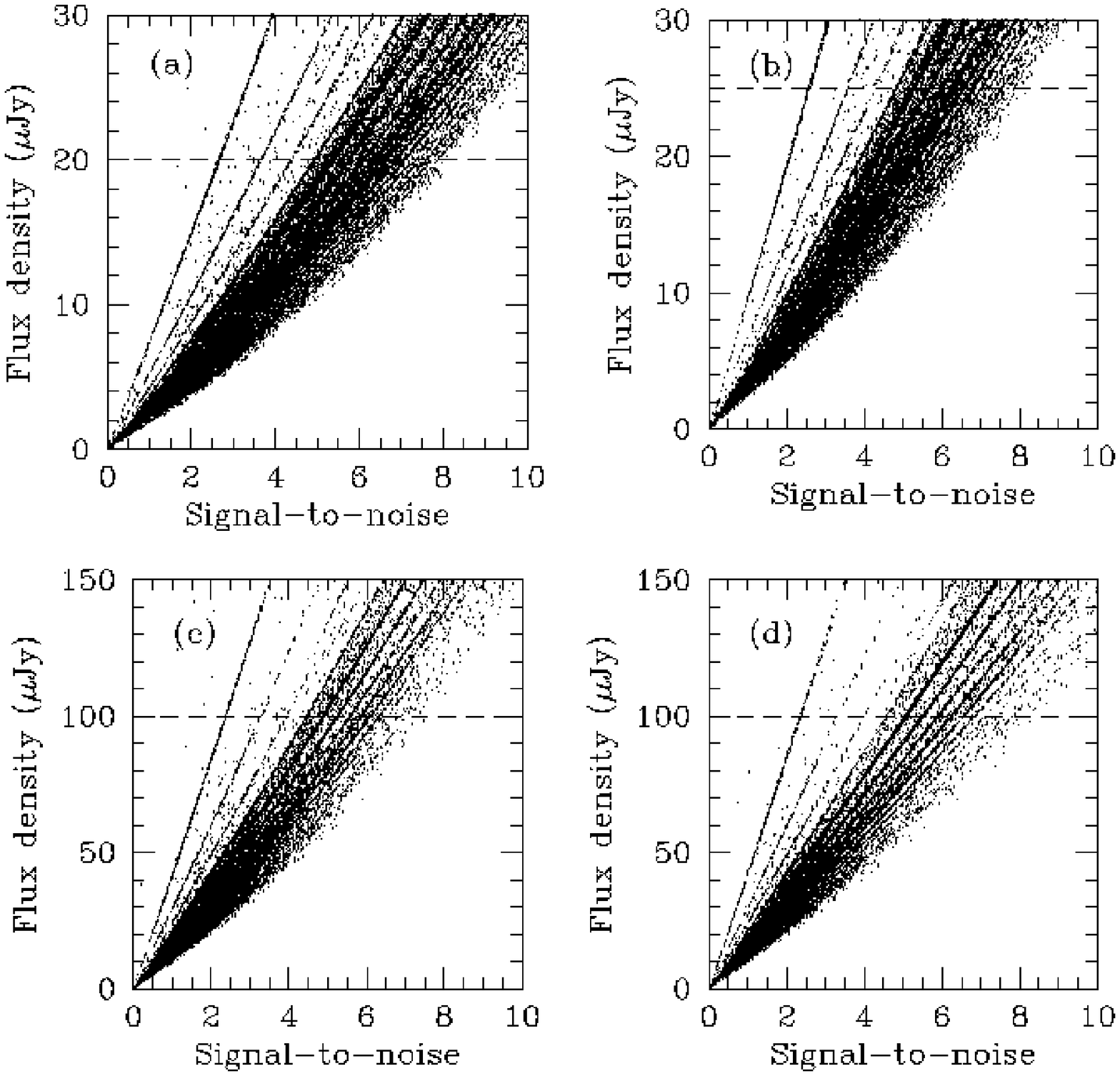}

\caption{Flux density against signal-to-noise ratio in the ``best'' 
apertures for sources in the raw main field catalogs
in each channel: (a) channel 1, (b) channel 2, (c) channel 3, and (d) 
channel 4. The dashed lines show the adopted flux density limits for the
final catalogs in each band. The striations in the plots are the result 
of selecting the fluxes from a small set of fixed apertures.}

\end{figure*}

\subsection{Catalogs}

\subsubsection{Single band catalogs}

The single-band catalogs contain the following fields: {\em srcID} - an
integer identifier for the source, {\em x, y} - pixel coordinates of the 
source in the mosaic, {\em RA, Dec} - right ascension and declination of the
source in the single-channel mosaic
(J2000 coordinates), {\em best flux} most reliable estimate of the 
flux density selected as described in Section 4, {\em best aper} - the 
aperture for the ``best'' flux (in arcsec), 
{\em err best flux} - the uncertainty in the 
``best'' flux, {\em S:N} - the signal-to-noise ratio in the ``best'' 
aperture, $6\farcs00 \, apc$ - aperture-corrected flux density in a $6\farcs00$
diameter aperture, $9\farcs26 \, apc$ - aperture-corrected flux 
density in a $9\farcs26$
diameter aperture, $14\farcs86 \,  apc$ - aperture-corrected flux density in a 
$14\farcs86$ diameter aperture, $24\farcs4 \, apc$ - flux density in 
a $24\farcs4$ diameter aperture (no correction applied as this is the 
aperture the IRAC calibration stars are measured in), $err \, 6\farcs00, 
err \, 9\farcs26, err \, 14\farcs86, err \, 24\farcs4$ - 
uncertainties in the aperture flux densities,
{\em area} - area of the isophotal aperture (in pixels), {\em Ixx, Iyy, Ixy} -
second moments of the flux distribution, {\em flag} - sum of the 
flag bits, see Table 3, {\em class}
- stellarity index from {\em Sextractor} (between 0 and 1, low values correspond
to probable extended sources and high values to point-like sources), 
{\em coverage} - number of unit frames contributing to the 
mosaic at the source position (see section 10.1). All fluxes and errors
are quoted in $\mu$Jy. The errors include the statistical error and 
the 10\% systematic uncertainty discussed in Section 7. The uncertainty 
in the overall flux density scale is not included, but we believe this to 
be small in any case, see section 10.3).
Table 7 summarizes the properties of the 
single-band catalogs. 

\begin{table}
\caption{Completeness and confusion statistics}

\begin{tabular}{lcrrrr}
Survey & Channel & Catalog & Completeness & Number &Source \\
       &         & limit   & at limit (\%)&of sources&density\\
       &         & ($\mu$Jy)&             &          & (deg$^{-2}$)\\\hline
Main          &1&20  &77&93689&24700  \\
Main          &2&25  &82&69915&18400\\
Main          &3&100 &93&10792&2840\\
Main          &4&100 &94&11891&3130\\
Verification  &1&10  &71&9339&37400\\
Verification  &2&10  &79&8652&34600\\
Verification  &3&30  &80&2904&11600\\
Verification  &4&30  &83&2517&10100\\
ELAIS-N1      &1&4   &75&2113&103100\\
ELAIS-N1      &2&3   &77&2381&116100\\
ELAIS-N1      &3&10  &90&1374& 67000\\
ELAIS-N1      &4&10  &90&919& 44800\\
\end{tabular}
\end{table}


\begin{table}
\caption{Numbers of sources in each catalogue as a function of band}
\begin{tabular}{lrrr}
detection flag & main & verification & ELAIS-N1\\
binary (decimal)&   &              &         \\\hline
0001 (1) &30758&2631&1271\\   
0010 (2) &6984&2061 &1622\\
0011 (3) &49534&3610&106\\
0100 (4) &822&139   &655\\
0101 (5) &698&666   &627\\
0110 (6) &24&16     &2\\
0111 (7) &2464&582  &34\\
1000 (8) &821&114   &297\\
1001 (9) &61&8      &1\\
1010 (10)&727&525   &543\\ 
1011 (11)&3510&368  &25\\
1100 (12)&103&12    &3\\
1101 (13)&11&2      &4\\
1110 (14)&23&18     &4\\
1111 (15)&6653&1472 &45\\
total in catalog&103193&12224&5239\\
\end{tabular}
\end{table}

\subsubsection{Four-band catalogs}

The four-band catalogs contain the following fields: {\em x,y} - pixel 
coordinates of the source in the shortest-wavelength mosaic in which it was 
detected, {\em RA, Dec} - right ascension and declination of the
source in the shortest-wavelength mosaic in which it was 
detected (J2000 coordinates), {\em ch1 id, ch2 id, ch3 id, ch4 id} - 
identifiers in the four single-band catalogs (defaults to -1 for 
non-detections in a given band), {\em ch1 flux, err ch1} - 
channel 1 flux density and uncertainty in the ``best'' aperture, {\em ch2 flux,
ch2 err} channel 2 flux density and uncertainty, {\em ch3 flux,
ch3 err} channel 3 flux density and uncertainty, {\em ch4 flux,
ch4 err} channel 4 flux density and uncertainty, {\em aper} - aperture used for 
flux density measurements, {\em flag} - see Table 4, {\em dflg} - sum of 
the detection
flag bits, see discussion in Section 4.

Objects with no coverage in a given band have their flux densities in 
that band set to 99999.0. Otherwise, 
non-detections have their fluxes set to the appropriate catalog limit, or
a 3-$\sigma$ limit, whichever is higher.

Table 8 shows the numbers of sources in each catalog as a function of 
detection flag. Of 93689 channel 1 main field detections, 62161 (66\%) 
are also detected in channel 2, 9826 (10\%) are detected in channel 3 
and 10235 (11\%) are detected in channel 4.

\section{Lessons Learned and Future Work}

The observations 
for the XFLS were designed well before launch, when the knowledge of the
properties
of the instrument was rather limited. 
With the benefit of hindsight we would have
changed a number of aspects of the design. The greater than expected scattered
light problems, and the latent image problems in channel 1 both 
could have been mitigated by using a larger dither pattern (with 
arcminute-scale dithers)
and/or half array offsets in the mapping strategy. On the other hand, 
one correct decision was
to maintain a relatively high coverage factor (five over the main field). 
This allowed reliable cosmic ray rejection, and the extra redundancy allowed
us to use the scattered light masks effectively. 

Future reprocessings of the XFLS data should be able to improve the accuracy
of the fluxes. For example, we may be able to use color-dependent flat 
fields to obtain more accurate fluxes by applying the 
flat field appropriate to the color of each object using an 
iterative technique. Also, we may be able to use improved
source extraction techniques based on a development of the SSC point source 
extractor to optimally-extract the typically 
slightly extended IRAC sources, and properly quantify the confusion in the
ELAIS-N1 channel 1 and 2 data. These improvements
were considered to be beyond the scope of this work, where our purpose is
to provide a reliable and fairly complete catalog 
on a reasonably short timescale.

\section{Summary}

The IRAC data from the XFLS has been analyzed, and a set of catalogs 
produced which we believe to be $\approx 80$\% complete and $\approx 99$\%
reliable. The final bandmerged catalogs contain 103193 objects in the 
main field, 12224 in the verification field and 5239 in ELAIS-N1. Flux 
densities of high signal-to-noise objects are accurate to about 10\%, and
the systematic uncertainty in the absolute flux 
density scale is $\sim 2-3$\%. Positional accuracy is $\approx 0 \farcs 25$
for high signal-to-noise objects and $\approx 1^{''}$ at the flux density
limits of the catalogs.
We have successfully extracted sources at source densities as high as 
100000 deg$^{-2}$ in our deepest channel 1 and 2 data, though there are 
indications that we are approaching the confusion limit at these high source
densities, in agreement with Fazio et al.\ (2004b). 
The mosaics and catalogs will be made available both through the 
{\em Spitzer} Science Archive and the 
NASA/IPAC Infrared Science Archive (IRSA). 

\acknowledgments

We thank the other members of the IRAC instrument/instrument
support team, in particular Sean Carey, for 
helpful discussions, and Dan Stern and Lexi Moustakas for making their
software available. We also thank I.\ Drozdovsky for supplying the 
mosaics of the ACS data in the XFLS field, and the 
anonymous referee for helpful
comments. This work is based on observations made with the {\em Spitzer Space
Telescope}, which is operated by the Jet Propulsion Laboratory, 
California Institute of Technology under NASA contract 1407.
The Two Micron All Sky Survey (2MASS) is a
joint project of the University of Massachusetts and the Infrared
Processing and Analysis Center/California Institute of Technology,
funded by the National Aeronautics and Space Administration (NASA) and the
National Science Foundation (NSF).

\end{document}